\documentstyle[12pt,aps]{revtex}
\tighten
\begin{document}
\title{{\Large {\bf 
Effective field theory at moderate finite density\bigskip}}}
\author{Boris Krippa\bigskip}
% \thanks{on leave from the 
%Institute for Nuclear Research of
% the Russian Academy of Sciences, Moscow Region 117312,
%Russia.}} 
%\vspace{1.5cm}
\address{ Department of Physics and Astronomy, Free University
of Amsterdam,\\ De Boelelaan 1081, 1081 HV Amsterdam.\\}
\maketitle
%\bigskip
%\vspace{-5.5cm}
\vspace{1cm}
\begin{abstract}
  Effective field theory of the in-medium nucleon-nucleon interaction
is considered. The effective range parameters are 
found to be of a natural scale. The low density limit is 
discussed both in perturbative and nonperturbative situations.
In the nonperturbative case the attractive character of the 
nucleon-nucleon interactions in the $^{1}S_0$ channel leads
to the nuclear superfluidity which is analyzed in the framework
of the renormalization group. The numerical values of 
the corresponding energy gap are in agreement with results
obtained by more more traditional many-body techniques.
The S-wave part of potential energy per particle is
calculated for different values of nuclear density.
The role of pion and many-body effects is discussed. 
Problems and challenges  in 
constructing the chiral theory of nuclear matter are outlined.   

\end{abstract}

\vskip0.9cm
%PACS nos.: 13.75.Cs, 11.10.Gh, 12.39.Fe, 11.55.Hx
%%\vfill
%\vskip0.2cm
%KEYWORDS: nucleon-nucleon interaction, effective Lagrangian, renormalization,
%nuclear matter, chiral symmetry
%\vfill

\newpage
{\bf 1. Introduction}.

Recently, there has been much interest in applying the
Effective Field Theory (EFT) \cite{Ma95} methods 
 to study nuclear interactions.
The key element of EFT is the counting rules  allowing for a consistent
 expansion scheme with predictable theoretical errors. Another important
 ingredient of any EFT is a separation of scales so that the high frequency
modes can be integrated out and manifest itself indirectly in the 
Low-Energy-Effective Constants (LEC's). In addition, the symmetries of the
underlying ``microscopic'' theory put important constrains on the 
interaction terms in the effective Lagrangian, making them less arbitrary
as compared to the sometimes more traditional phenomenological description.
In the context of nuclear matter such a description implies the use of 
some phenomenological potentials  leading, in general, to the different
 effective in-medium interactions and bringing uncontrollable uncertainties,
related to the off-shell ambiguities. These ambiguities can, at least
in principle, be eliminated in the EFT approach by using the freedom to
work in the different low-energy representations of the underlying QCD.
 In the other words EFT  may be useful to get rid of the long-standing
problem arising in the more standard treatments of  hadron
interactions and related to the necessity of utilizing of the phenomenological 
form-factors to regulate the divergent integrals and parameterize the
 complicated internal structure of hadrons. There is nothing wrong to use them
in the calculations based on phenomenological models. However, the
 problem is that the underlying field theory must then be nonlocal making
the whole issue of constructing the entirely consistent field-theoretical
treatment of the low-energy hadron interactions extremely difficult 
to deal with. Moreover, when the 
problem under consideration is the interaction of hadrons with the external
electromagnetic fields, 
 the straightforward use of form-factors may lead
to the conflict with gauge invariance.
 All these features make the reanalysis of nuclear matter
in the framework of EFT useful and promising. The main element of nuclear
 dynamics is the elementary nucleon-nucleon interactions. Because of the 
large scale involved in this process
the EFT treatment of the NN-interactions in vacuum must be
nonperturbative. In the framework of EFT the problem was first considered 
by Weinberg \cite{We} who suggested to apply chiral counting rules to the
set of all possible irreducible diagrams at given order which  are then  to
be iterated in all order by solving Lippmann-Schwinger
equation. In somewhat different approach  proposed
 by Kaplan, Savage and Wise (KSW) \cite{Ka}  only certain
 subclass of lowest order diagrams
was summed up to all orders whereas the rest was treated as a perturbation
so that KSW counting is applied directly to the scattering amplitude.
The analysis \cite {Bi} of the NN (pionless!) scattering amplitude
 based on the Wilson renormalization group has demonstrated that 
power counting indeed closely follows that which was found by KSW. 

The extension of these developments to the case of finite density is by
no means straightforward \cite{Kr}. There are several points which must be 
taken into consideration to formulate the consistent EFT at finite baryon 
density. First, being unnaturally large in vacuum the NN 
S-wave scattering lengths get significantly reduced in nuclear matter
due to Pauli principle and many-body interactions. Second, the way of 
formulating the counting rules allowing for the 
consistent treatment of  the many-body effects, such as the ring and ladder
diagrams must be found. Third, the long-range part of the NN
effective interaction related to the pion exchange in nuclear environment 
should be taken into account. In addition, the three and four
 nucleon interaction terms must be included in the EFT Lagrangian.
 
Some recent studies of of the systems with the finite fermion density in
 the context of EFT have been focused on constructing of the chiral 
expansion for the dilute interacting Fermi-gas \cite{Fu1} and mean-field
effective chiral Lagrangian \cite{Se}. The issues of the
 in-medium counting rules 
\cite{Lu} and nuclear binding in chiral limit \cite{Mil} were also examined 
recently. In this paper we focus on the EFT approach to construct the analog
of the Brueckner G-matrix and try to relate 
the in-medium  LEC's and those obtained
in the free space fit. It seems to be quite a nontrivial problem to establish
 such   relations since the LEC's entering the effective potential of the 
in-medium NN interactions may be density dependent.

Any EFT expansion has a form in $k/\Lambda$, the ratio of 
some generic momentum involved and  a characteristic short range physics mass scale.
So it is important to establish a  relevant expansion parameter. The analysis
of the different phenomenological models indicates that the 
nonrelativistic mean field
approximations, successfully describing the bulk nuclear properties at low and
 moderate energy scale, encounter difficulties when applied to the 
nuclear phenomena with a characteristic mass scale around 600 MeV 
(electromagnetic form-factors at relatively large momentum 
transfer, for instance). This is the region where the effect of the short range
correlations (SRC) becomes important. Thus, it is plausible that the typical scale
where the EFT description of nuclear matter with pointlike nucleon-nucleon interaction
can be close to the inverse range of the SRC.
One notes that the addition of the long-ranged pion effects should
not significantly change this estimate. As the typical nuclear
momentum scale is of order the Fermi-momentum $p_F \simeq$ 260-270 MeV the relevant 
expansion parameter is close to 0.3 - 0.4. This  admittedly crude estimate
can be improved if the influence of the nuclear many-body environment is taken into
account. 

All essential features of the  nucleon-nucleon interactions can, in principle, be obtained
from the effective chiral Lagrangian  
 \begin{equation}
{\cal L}=N^\dagger i \partial_t N - N^\dagger \frac{\nabla^2}{2 M} N
- \frac{1}{2} C_0 (N^\dagger N)^2\\ 
-\frac{1}{2} C_2 (N^\dagger \nabla^2 N) (N^\dagger N) + h.c. + 
\label{eq:lag}
\end{equation}      
We consider the simplest case of the NN scattering in the $^{1}S_0$ state
 and assume
zero total 3-momentum of NN pair in the medium. The inclusion of
the nonzero total 3-momentum does not really change anything 
qualitatively and only makes the calculations technically more involved.   
The G-matrix, describing the dynamics of the in-medium interactions satisfies
the BG equation
  
The G-matrix is given by
\begin{equation}
G(p',p)=V(p',p) + M \int \frac{dq q^2}{2 \pi^2} \, V(p',q) 
\frac{\theta(q-p_F)}{M(\epsilon_{1}(p) +\epsilon_{2}(p')) - q^2} G(q,p),
\label{eq:LSE2}
\end{equation}

Here $\epsilon_1$ and $\epsilon_2$ are the single-particle 
energies of the bound nucleons.  In nuclear medium 
 bound nucleon acquires the  effective mass slightly different from that in free space.
In a fully self-consistent calculations the nuclear mean field 
and chiral G-matrix should, in principle,
 be treated using the same Lagrangian. However, it requires the inclusion 
both pion and many-body effects and establishing the power counting rules at finite density.
In this paper we explore an easier approach and use the in-medium value of the
nucleon mass close to the accepted one in the phenomenological
nonrelativistic \cite{Ma} mean field treatments.
We used the value $M = 0.75 M_0$ in the calculations at $p_F$ = 1.36 $fm^{-1}$,
 where we denote $M_0$  the nucleon mass in vacuum.
In order to check the sensitivity of the results obtained to the different choices of
effective nucleon mass
the values of $M$ have been varied from $0.7 M_0$ to $0.8 M_0$. The 
higher order changes of the results were found  for the potential energy per particle
so we will keep the value of $M$ fixed and discuss the results corresponding the choice
$M = 0.75 M_0$.

This paper is organized as follows. In the next section we consider an exactly
solvable model to estimate the in-medium effective range parameters. Then 
the generalization of the effective nucleon-nucleon Lagrangian to the 
finite density case is discussed.
 In the  section 4 we consider the
 low-density limit in the case when the perturbative treatment is possible.
The issue of superfluid nuclear gap is discussed in the  section 5.
Section 6  offers a brief summary  of results for a potential energy 
of nuclear matter and some concluding remarks.
 
{\bf 2. Exactly solvable model.}

In order to get an idea about the typical size of the
in-medium  effective range parameters which, in turn, determine the 
typical scale of the problem it seems reasonable to exploit some exactly
solvable model with the parameters adjusted to describe the experimental data
or empirical results. Moreover, the in-medium scattering amplitude obtained
in such solvable model can further be used to extract the values of the
 effective couplings  in the corresponding EFT. In some sense this is similar
 to the procedure usually adopted in the vacuum case. Firstly, one should
 determine the LEC's from, say, phase shifts and then use these LEC's
to calculate the other observables. Unfortunately, there is no such thing as
phase shift analysis in nuclear matter.
  One could instead  rely on some solvable 
model with the parameters adjusted to describe nuclear data  
to extract the numerical values of the effective couplings. However,
there is an important difference between vacuum and in-medium cases.
In the former the corresponding LEC's  are extracted from the model 
independent experimental data whereas in the latter the in-medium LEC's are
determined from the model dependent scattering amplitude  which may, in
general, differ for the phase-equivalent NN potentials. In this respect it
would be desirable to relate the vacuum and in-medium effective couplings
to be able to use the experimental data on the NN scattering in vacuum.  
We will discuss this and related issues in more details below.

We use a model with a simple separable interaction    
\begin{equation}
V=-\lambda |\eta\rangle\langle\eta|
\end{equation}
with the form factors

\begin{equation}
\eta(p)=\frac{1}{(p^2 + \beta^2)^{1/2}}
\end{equation}
 One can easily get 
\begin{equation}
\frac{1}{T(k,k)}=
{V(k,k)^{-1}}\left[1 - M_0 \int \frac{dqq^2}{4 \pi^{2}} \, \frac{V(q,q)}
{{k^2}- q^2}\right] 
\label{eq:sep}
\end{equation}
The analytic solution for the corresponding $T$-matrix is straightforward
\begin{equation}
\frac{1}{T(k,k)}=
{V(k,k)^{-1}}\left[1 - M_0 \int \frac{dqq^2}{4 \pi^{2}} \, \frac{V(q,q)}
{{k^2}- q^2}\right] 
\label{eq:sep}
\end{equation}
The G-matrix being the solution of the BG equation for the separable 
interaction takes the form 
\begin{equation}
G(k,k)=-
\eta^{2}(k)\left[\lambda^{-1}+ 
\frac{M}{2 \pi^{2}} \int{dqq^2} \, \frac{\theta(q-p_F) \eta^{2}(q)}{{k^2}- q^2}\right]^{-1}
\end{equation}
The empirical value of the potential energy per particle is $\sim$ -16 MeV
and can be reproduced with 
\begin{equation}
\lambda = 1.95\,\qquad \beta = 0.8\, {\rm fm}
\end{equation}
These parameters being 
substituted to the G-matrix in the low-momentum limit $G(k \rightarrow 0)$
lead to the estimates 
$a_m\simeq r_m\simeq 0(1)$, where $a_m$ and
  $r_m$ are the in-medium analogs of scattering length and effective radius.
So one can conclude that the in-medium effective range parameters are of
 a natural size which is roughly 
given by the value of the Fermi-momentum
at nuclear saturation $ p_{F} \simeq 1.36fm^{-1}$. This allows us to avoid
 some part of the difficulties typical for the EFT for the
nucleon-nucleon interaction in vacuum  such as necessity of a 
special treatment for the system with the unnaturally large scattering length.
Of course, there are many other specific points  typical 
 for the finite density EFT and making its consistent yet practical
 realization quite nontrivial. Some of them will be considered in what
 follows. As already mentioned above the in-medium phenomenological amplitude
 may depend on the model of the NN interaction used. To check whether
the freedom to choose among phenomenologically equivalent
models of the NN interactions may give rise to the
 significant changes of our estimates of the in-medium effective
range parameters we carried out the calculations with the
exponential parameterization of the form-factors entering
the  separable potential. The values of   $a_m$ and $r_m$ found are as well 
of a natural size $\simeq 0(1)$ and its numerical values are close to those
obtained with the ``square root'' fit. So it seems that the conclusion
about a natural size of the the in-medium effective range parameters 
is quite robust and practically model independent. Physically it looks quite 
natural. Firstly, the integrand in the BG equation does not have a pole
irrespectively of the explicit form of the kernel. Secondly, the strength
of the shallow virtual nucleon-nucleon bound state gets significantly 
reduced in nuclear medium because of interaction with nuclear mean field
leading to the inevitable decrease of the scattering length.

{\bf 3. Effective Lagrangian treatment.}

The solution of the BG equation with the NN potential extracted from the 
effective Lagrangian is similar to that in the vacuum case \cite {Co}
and given by
\begin{equation}
\frac{1}{G(k,p_F)}=\frac{(C_2 I_3(p_F) -1)^2}{C_0 + C_2^2 I_5(p_F)
 + {k^2} C_2 (2 - C_2 I_3(p_F))} - I(k,p_F),
\label{eq:Tonexp}
\end{equation}
where we the loop integrals are

\begin{equation}
I_n (p_F) \equiv -\frac{M}{(2 \pi)^2} \int dq q^{n-1}\theta(q-p_F).
\label{In}
\end{equation}
 
and  

\begin{equation}
I(k,p_F) \equiv \frac{M}{2 \pi^{2}}\int dq  \, \frac{q^2\theta(q-p_F)}
{{k^2}- {q^2}}.\label{eq:IEdef}
\end{equation}
These loop integrals are divergent and therefore the procedure of 
regularization and renormalization must be carried out. Note that the
 issue of the nonperturbative renormalization is quite a subtle problem.
In contrast to the standard perturbative case where the usual field
 theoretical methods can be used to regularize the given divergent graphs and then  
renormalize the bare coupling constants, in the nonperturbative situation
the renormalization of the whole integral equation must be carried out.
In the case when the analytic solution for the scattering amplitude can be obtained
as, for example, in the pionless nucleon-nucleon EFT, the renormalization of the
 amplitude is a rather straightforward procedure. However, if the explicit solution
 is not possible (this is the case for the realistic NN forces) then the special
 care is needed to perform the renormalization in a consistent way \cite {KrB}.
In this paper we follow the procedure used in Ref. \cite {Ge} to renormalize the
effective NN amplitude in vacuum. 
 We subtract the divergent integrals at
some kinematical point $p^2 = -\mu^2$. After subtraction the renormalized G-matrix
 takes the form

\begin{equation}
\frac{1}{G(k,p_F)}=\frac{1}{C_{0}(\mu, p_F) 
 +  2{k^2} C_{2}(\mu, p_F)} +\frac{M}{4\pi}[ k\ln\frac{p_F - k}{p_F + k}
 - i\mu\ln\frac{p_F - i\mu}{p_F + i\mu}] ,
\label{eq:GR}
\end{equation}
where the couplings $C_{0(2)}$ should now be interpreted as the renormalized quantities
depending on some renormalization scale $\mu$.
It is easy to see that in the 
 $p_F \rightarrow 0$ limit  the vacuum chiral NN amplitude is recovered.
 The $\mu$ dependence of LEC's is 
governed by the renormalization group (RG) equations. We require 
 G-matrix to be a subtraction point independent quantity.
 Applying  $\partial /\partial \mu$ to the expression for the G-matrix
 and setting $\partial G/\partial \mu$ = 0 one can get the following RG equations
\begin{equation}
\frac{\partial C_0(\mu, p_F)}{\partial \mu} = \frac{C_0^2 M}{2\pi^2}(\frac{\mu p_F}{p_F^2 + \mu^2}
+  arctan(\frac{\mu}{p_F}))
\end{equation} 
\begin{equation}
\frac{\partial C_2(\mu, p_F)}{\partial \mu} = \frac{C_0 C^2 M}{\pi^2}(\frac{\mu p_F}{p_F^2 + \mu^2}
+  arctan(\frac{\mu}{p_F})).
\end{equation} 
In the limit $p_F \rightarrow$ 0 these equations transform to the ones derived 
by Kaplan et al. \cite{Ka}. The solutions of these RG equations are
\begin{equation}
C_{0}(\mu, p_F) = \frac{C_{0}(\mu_{0}, p_F)}{1 + 
\frac{M}{2\pi^2}(\mu_0 arctan(\frac{\mu_0}{p_F}) - 
\mu arctan(\frac{\mu}{p_F}))C_{0}(\mu_{0}, p_F)} 
\end{equation}
and
\begin{equation}
C_{2}(\mu, p_F) = 
C_{2}(\mu_0, p_F)  \left[\frac{C_{0}(\mu, p_F)}{C_{0}(\mu_{0}, p_F)}\right]^2 
\end{equation}
In order to determine the scale dependence of the effective couplings one 
needs to define its boundary values at some kinematical points. 
To extract these values we equate the EFT and phenomenological
 expressions for the G-matrix at $p=p_F/2$ and  $p=p_F/3$. This is 
somewhat similar to the procedure used to get the values of the vacuum LEC's
when the fit is usually done using the experimental phase shifts
within some kinematical region.
One notes that when $p_F \rightarrow$ 0 the above expressions
for the effective couplings 
reduce to these obtained in Ref. \cite{Ka}
The assumed value of the Fermi-momentum is $p_F=1.36 fm^{-1}$.
If the value $\mu = 0$ is chosen as a subtracting point we find
 $ C_{0}(\mu=0,p_F) = -1.88 fm^{2}$ in LO. For the lowest order
coupling fixed at NLO we get  
$ C_{0}(\mu=0,p_F) = -2.67 fm^{2}$ and $C_{2}(\mu=0,p_F)= 0.85 fm^{4}$.
Substituting these values in the RG equations we get
 $C_{0}(\mu=m_{\pi},p_F) = 2.35 fm^{2}$ and $C_{2}(\mu=m_{\pi},p_F)= 0.64 fm^{4}$.
The obtained effective  coupling $C_0$ at $\mu = m_{\pi}$ is fairly close to its
 vacuum value \cite{Ka}. Thus one can conclude that the density dependence
 of the in-medium LO couplings is rather moderate, provided that 
 $\mu \simeq p_F$. It means that at densities smaller than the normal
 nuclear one there is a possibility to use the vacuum values of the
 corresponding LEC's to get an idea about the approximate order of
their in-medium analogs. It seems that such a matching of the vacuum and in-medium
LEC's could provide  a quite important additional
possibility allowing one to put some constraints on the values of the in-medium LEC's
since the vast experimental data on the NN scattering in vacuum can be used
in the calculations at finite (although moderate) densities. It may also 
turn out useful when calculating the nuclear saturation curve since   
it helps to avoid the procedure of fixing the phenomenological 
in-medium amplitudes at different nuclear densities which would otherwise
be needed to fix the density dependence of the effective couplings,
making the whole EFT approach less predictable and reliable.
Note that, although this matching can be carried out at any value of $\mu$
the choice $\mu = 0$ is rather inconvenient one since
 the values of the vacuum and in-medium LEC's
compared at $\mu = 0$ are completely different so that the conclusion about
the moderate density dependence of the effective couplings may no longer
 be true at  $\mu = 0$. Thus, the choice   $\mu \simeq m_{\pi} \simeq p_F$ looks more
natural. We emphasize that the observables, expressed via G-matrix do
 not depend on the value of  $\mu$ chosen. However, it is much more difficult 
to relate the in-medium and vacuum NN-forces when the value  $\mu = 0$
is taken for the subtraction point.

We observe approximately 35$\%$ change
in the value of  $C_0(\mu = m_\pi, p_F)$ when going from LO to NLO. It 
indicates that the chiral expansion is systematic in a sense that
  the NLO corrections lead to the ``NLO changes'' of the effective couplings already determined at
LO. 
 The natural size of the in-medium effective range parameters, moderate changes
experienced by the LO coupling constant $C_0$ and smallness
of the NLO LEC's might, in principle, indicate the possibility
of the perturbative calculations.   However, in spite of
this, it is still more
useful to treat this problem in the nonperturbative manner. There are few 
reasons for this. Firstly, the overall (although distant)
goal of the EFT description is to derive both nuclear matter and the 
vacuum NN amplitude from the same Lagrangian. However, it is hard to say 
at what densities the dynamics becomes intrinsically nonperturbative, so it is 
better to treat the problem nonperturbatively from the beginning.
The nonperturbative treatment may also turn out important to get the 
correct saturation curve since at some density lower than the normal
 nuclear one the scattering length starts departing from its natural value
and some sort of the nonperturbative approach becomes inevitable.
Besides, the logarithmic terms in the chiral G-matrix are important to 
ensure correct threshold behavior of the in-medium NN-amplitude. 
Secondly, the NLO corrections themselves
are quite significant.
Thirdly, the nonperturbative treatment is required to study the phenomena
of nuclear superfluidity \cite {Mi} when the perturbation theory is not valid.
Finally, in the processes
involving both the nonzero density and temperature, such as heavy ion
collisions, the value of the Fermi-momentum can effectively be lowered again
making the nonperturbative treatment preferable. 

As we mentioned in the introduction , one of the most important  (and yet unsolved)
problems of constructing EFT at finite density is the formulation of the corresponding
counting rules. The complete solution of this problem is possible only if pion effects
and many-body forces are taken into account in a chirally invariant manner. 
 However, as a first step in this direction one could
 formulate the ``naive'' counting rules  for the LEC's $C_{0(2)}$ used in
 the effective Lagrangian.
Since the main reason of the anomalous KSW counting 
$C_{2n} \sim \frac{4\pi}{M \Lambda^{n} \mu^{n+1}}$ suggested by Kaplan et al. \cite {Ka} is no
 longer the case in nuclear medium it would be natural to count like  
$C_{2n} \sim \frac{4\pi}{M \Lambda^{2n+1}}$. However, one could modify this counting 
taking into account a further small scale $Q \sim p_F \sim m_\pi \sim \mu$
 involved at nonzero density
so that taking $p_F \rightarrow 0$ limit gives rise to the KSW scheme. So we count
\begin{equation} 
C_{2n} \sim \frac{4\pi}{M \Lambda^{n}\mu^{n+1}}
\left[ \frac{\mu+p_F}{-\mu - (p_F / \mu)(\Lambda -\mu)} \right]^{n+1}.
\end{equation} 
 Assuming $\mu \sim p_F$ one indeed observes the
 moderate dependence of the LEC's on density. 
The above stated  values of LEC's are consistent with this power counting.
 Note that in order to be fully 
consistent the density dependent nucleon mass should, in principle, be used
 in Eq.(17).
However, to get the crude estimate of the effective couplings the use of vacuum
 value of nucleon mass
is sufficient. It is clear that the exact form of the density dependence of the effective
couplings is not unique. Any functional form providing the smooth interpolation between the 
``natural'' counting $C_{2n} \sim \frac{4\pi}{M \Lambda^{2n+1}}$ at normal nuclear
density and the KSW counting $C_{2n} \sim \frac{4\pi}{M \Lambda^{n} \mu^{n+1}}$ at zero  density
is acceptable. It is plausible however that all forms will give the same order-of-magnitude
estimate of the effective coupling. When calculating some observables at normal nuclear density
the simplest thing is to make use of the ``natural counting''. The further understanding of
the density dependence of the in-medium LEC's could be achieved considering the nuclear saturation
since the saturation curve would be quite sensitive to the exact values of the LEC's at 
different densities. However, to treat the saturation in a consistent manner the higher partial
waves (at least P and D waves) and pion effects must be included. We relegate the discussion
of nuclear saturation to the separate paper. Note that although the value of the in-medium 
cutoff parameter $\Lambda$ may also differ from its vacuum value we do not 
expect this difference to be significant. It seems to be rather unlikely
that the introduction of the additional low-momentum scale $p_F$ can lead to the large
 changes of
the  cutoff $\Lambda$, which is meant to be the indirect manifestation of the truly
 short-range physics which is only moderately touched upon by the medium effects.
 
 Assuming that the values of the effective couplings are fixed at
 zero  nuclear density and using the above written expression for the couplings
as the functions of density one can determine the LEC's at any intermediate
value of the Fermi-momentum. 

Now the remark, concerning the naturalness criteria \cite{GeM} is in order.
The naturalness concept in the context of nuclear interactions was elaborated
in   \cite{Se}.
According to this concept of  an individual term in
the effective Lagrangian can schematically be written as some dimensionless factor
of order unity multiplied by the certain combination of scale factors,
characterizing the scales involved 
\begin{equation}
C \sim c\left[\frac{\psi^{+}\psi}{f_{\pi}^2\Lambda}\right]^l
\left[\frac{\partial}{\Lambda}\right]^n (f_\pi\Lambda)^2.
\end{equation}  
Applying the scaling rules developed in \cite{Se} to extract
 all scale factors and assuming that the cutoff 
parameter $\Lambda \simeq$ 500 MeV
one finds $c_0(c_2) \simeq 0(1)$. Thus the 
dimensionless coefficients are indeed compatible with naturalness.

 {\bf 4. Low-density limit. Perturbative case}

The low density limit has often served as a somewhat toy many-body problem, useful for
 understanding the structure of the theory and checking the consistency of the
 approximations made.
In this section we consider the case when all interactions are natural so that the perturbative
treatment is valid. The corresponding expression for the potential energy per particle 
of the dilute Fermi-gas is well known and was derived by traditional many-body methods long
time ago. We show in this section that in the limit when the perturbative expansion is
valid $(G \sim V)$ the LO term of the low-density expansion can be reconstructed within
 our approach.
Note that the low-density expansion including both leading and higher order terms has recently
been rederived in ref. \cite{Fu1} in the framework of perturbative EFT.
 The pairing phenomena requiring the nonperturbative treatment will be considered in the next section. 

 The potential energy
of the system of the interacting fermions is
 \begin{equation}
U_{tot} = \frac{1}{2}\sum_{\mu,\nu}<\mu\nu|G(\epsilon_\mu + \epsilon_\nu)
|\mu\nu-\nu\mu)
\end{equation}
The summation goes over the states with momenta below $p_F$.
Since by assumption the perturbative expansion is valid we put $(G \sim V \sim C_0)$.
The explicit expression for the  $S$-wave part of  ground-state energy per particle is
\begin{eqnarray}
\frac{E}{A} =\frac{3 p_{F}^{2}}{10 M_0} + 
 \frac{3}{\pi^2 p_{F}}(2T+1)\int_{0}^{p_F} dk k^2 G(k,p_F)\nonumber\\
\left[\int_{0}^{p_{F}-k} dP P^2 
+ \int_{p_{F}-p}^{(p_F-p^2)^{1/2}} dP P^2 \frac{p_F - P^2 -k^2}{2Pk}\right],
\end{eqnarray}
where $P$ is the total momentum of the pair.
Since we consider the low-density limit it is legitimate to use the relation
$C_0 = \frac{4\pi a}{M_0}$
 between the effective coupling $C_0$ and scattering length $a$. After the
 momentum integration is
done we obtain the following lowest order expression for the ratio $\frac{E}{A}$ in the case
of the dilute Fermi-system 
\begin{equation}
\frac{E}{A} = \frac{p_{F}^{2}}{M_0}\left[\frac{3}{10} + \frac{1}{3\pi}(p_F a) + O(p^{2}_F) \right],
\end{equation}
which coincides with the well known textbook expression \cite{FeW}. The next analytic
in $p_F$ terms of the low-density expansion can be derived if NLO term with the effective
coupling $C_2$ is included. The result also agrees with that obtained by standard many-body
 technique. One notes that it may be nontrivial to go further in the low-density expansion
since some higher order terms appear in the conventional expression multiplied by powers of 
$ln(p_{F}a)$. As argued in Ref. \cite{Fu1} in order to reproduce these terms in the 
framework of the perturbative EFT the graphs describing three-to-three scattering  amplitude
must be included. The point is that the diagram for the two-to-two scattering contain
 only power divergences and thus, being regularized by means of dimensional regularization
with minimal subtraction, can only lead to the terms which are analytic in $p_F/\Lambda$.
Note that it may not be so in the nonperturbative situation with unnatural scattering length
since the G-matrix from Eq.(12) already contains the logarithms of $p_F$ needed to 
reproduce the terms with $ln(p_F a)$ in the low-density expansion so that no additional
3-body terms are needed. Of course, in the case of large scattering length the low-density
expansion has a little practical use since it is valid for very low values of $p_F$,
satisfying the condition  $p_{F}a < 1$ 

{\bf 5. Low-density limit. Nonperturbative case}

It is well known that nuclear matter is superfluid. The nucleon-nucleon
 interactions is attractive in the $^{1}S_0$ channel so that the corresponding
amplitude develops the singularity, no matter how weak the interaction is.
This singularity  is cured by the formation 
of Cooper pairs. The superfluid properties of nuclear matter are important
in the study of neutron stars \cite{Pe} and heavy nuclei close to the drip
line \cite{Mu}. The studies of nuclear superfluidity  have basically been
carried out using pairing matrix element given by the bare  
 nucleon-nucleon interactions as the elementary input entering the many-body 
calculations so it would be useful to see how nuclear superfluidity is described when the
 basic nucleon-nucleon interaction is treated in the framework of EFT.

The gap exhibited by the system  is formed at the Fermi surface. Since Cooper
pair condensation cannot be described by means of a perturbation
 theory a nonperturbative treatment is required. The standard approaches 
to the problem of nuclear superfluidity are based on the traditional
many-body methods such as coupled-cluster theory \cite{Bi} or 
self-consistent Green function theory \cite{Di}. To our knowledge the first
attempts to analyze the nuclear superfluidity in the framework of EFT was 
made in Refs. \cite{Ho,Pa}. The discussion below follows the line which is
 somewhat similar to that in Ref. \cite{Ho}. The aim of this section is 
to use the RG equations to estimate the size of the superfluid gap at small
nuclear densities. In the RG technique the size of the gap is determined
by the position of the singularity in a running effective coupling near the 
Fermi surface \cite{Po}. Let's define
\begin{equation}
\frac{\mu_0}{p_F} arctan(\frac{\mu_0}{p_F}) = t_0 ,  
\frac{\mu}{p_F} arctan(\frac{\mu}{p_F})) = t. 
\end{equation}   
After substitution to Eq.(16) one gets
\begin{equation}
C_{0}(t, p_F) = \frac{C_{0}(t_{0}, p_F) 2\pi^2}{2\pi^2 + 
 M p_F (t_0  - t)C_{0}(t_{0}, p_F)}. 
\end{equation}
We take the limit $p_F \rightarrow 0$ so that  the 
effective coupling $C_0$ extracted from the nucleon-nucleon
 scattering in vacuum can be used
\begin{equation}
C_{0}(\mu_0, p_F \rightarrow 0 ) \simeq  C_{0}(\mu_{0}) 
= \frac{4\pi}{M_0}(\frac{1}{-\mu_0 + 1/a}). 
\end{equation}
We denote $a$ the vacuum scattering length. One can see that the 
coupling $C_0$ has singularity at 
\begin{equation}
\nu^* =  \frac{4\pi}{M_{0} C_{0}(t)} +\frac{2 p_F}{\pi} t.
\end{equation}
The pairing gap is expected to be of size $\sim exp(\frac{\pi}{2 p_F}\nu^*)$.
Assuming $t = 0$ one can rewrite the expression for the gap in terms 
of the scattering length 
\begin{equation}
\Delta  \sim  b exp(\frac{\pi}{2 p_F a}). 
\end{equation}
This expression coincides with one obtained in Ref.\cite{Pa}.
However, as pointed out in \cite{Ka} the choice $\mu = 0$ is not 
the optimal one. The choice $\mu \sim p$ makes the power counting
more transparent. Moreover, at $\mu \sim 0 $ the density dependence
of the $C_{0}$ may become nonnegligible even at low densities so 
the assumption $C_{0}(\mu, p_F) \simeq C_{0}(\mu, p_F=0)$ is better justified
when  $\mu > p_F$.
The scattering amplitude is independent of the subtraction point so the 
choice of this point is just a matter of practical convenience.
We used the value $C_{0}(\mu) = -2.23 fm^{2}$ determined at  
$\mu = 1.2 fm^{-1}$ and calculated the gap $\Delta (p_F)$ as a function
of Fermi momentum. One notes that the RG analysis gives the order of magnitude
estimate of the gap but cannot provide the determination of the preexponential
factor which may be important at small $p_F$. This factor can be found by
 matching to the known results, obtained at low density \cite{FeW}.
The result of this matching is 
\begin{equation}
b \simeq  \frac{p_{F}^2}{2 M}. 
\end{equation}
The results of calculations of $\Delta (p_F)$ are
\begin{eqnarray}
\Delta (p_F = 0.1 fm^{-1}) = 0.075 MeV, 
 \Delta (p_F = 0.2 fm^{-1}) = 0.3 MeV,\nonumber\\
 \Delta (p_F = 0.4 fm^{-1}) = 0.76 MeV,
\Delta (p_F = 0.6 fm^{-1}) = 2.39 MeV,
\end{eqnarray}
We stopped calculations of the gap at $p_F = 0.6 fm^{-1}$ since at higher values
of Fermi momentum the condition $\mu > p_F$ is no longer valid and the 
density dependence of the effective couplings may affect the results of 
calculations.

Our results are in a qualitative agreement with the calculations using more 
traditional approaches. However, as well known 
from the phenomenological studies at $p_F \simeq 1.7 fm^{-1}$ the gap vanishes.
The physical reason is that at such densities the attraction in the $^{1}S_0$
 channel is replaced by the repulsion so that the formation of the fermion
condensate is no longer possible. The EFT treatment must be modified
to incorporate the effect of the decrease of a  superfluid gap with
increasing density since the ration $p_F/\Lambda$ is not a reliable
expansion parameter any longer and density dependence of the LEC's should be
 taken into account. The problem of the EFT description of nuclear superfluid
gap in the wider range of densities will be addressed in a separate paper.

{\bf 6. Nuclear matter observables}   

Let's now discuss the results of calculations of potential energy of
nuclear matter. As in the low-density case the Eq.(20) is used. 
Let's start from the results of calculations at normal nuclear density
$\rho_0 = 1.36 fm^{-1}$. In the LO perturbative calculation assuming 
$G \sim C_0$ one gets $\frac{U(^1S_0)}{A}\simeq -11.9 MeV$. Making use of the
 nonperturbative expression for the G - matrix but keeping only LO in the
 chiral expansion of the effective nucleon-nucleon potential gives rise
to the value  $\frac{U(^1S_0)}{A}\simeq -17.0 MeV$. The inclusion of the NLO
terms in the chiral Lagrangian leads to  $\frac{U(^1S_0)}{A}\simeq -13.0 MeV$.
So the corresponding correction is about 30$\%$ and can be attributed to 
NLO. One notes that both  $\frac{U(^1S_0)}{A}$ and LO effective coupling $C_0$
 experience NLO corrections when NLO terms are included in the effective
Lagrangian. It justifies one more time the statement about the consistency
of the corresponding chiral expansion. The similar calculations performed
in the triplet S-wave channel leads to the value
 $\frac{U(^3S_1)}{A}\simeq -17.3 (-13.2) MeV$ in LO (NLO). However, there is 
an important difference between the singlet and triplet channels because of
the S-D mixing arising at NLO in the triplet channel. This effect is not
 considered in this paper since it requires the inclusion of the explicit
pion degrees of freedom  in the Lagrangian. The issue of the pionic effects
is discussed in more details below. Here we only note that the pion
 effects must eventually be included in a chirally symmetric manner
consistent with the in-medium counting rules that are yet to be established. It makes
 the whole issue of the pionic effects quite complicated.
 To calculate the value of 
 $\frac{U}{A}$ at lower densities we use Eq.(17) to fix the effective 
couplings $C_0$ and $C_2$. We use the value $\Lambda = 2.05 fm^{-1}$ for the
 effective cutoff parameter. The effective couplings determined at
different densities and at $\mu = m_{\pi}$ are 

\begin{eqnarray}
C_{0}(\mu, p_F = 1.2 fm^{-1}) = -2.41 fm^2 ;
C_{2}(\mu, p_F = 1.2 fm^{-1}) = 0.98 fm^4\nonumber\\
C_{0}(\mu, p_F = 1.0 fm^{-1}) = -2.48 fm^2 ;
C_{2}(\mu, p_F = 1.0 fm^{-1}) = 1.05 fm^4\nonumber\\
C_{0}(\mu, p_F = 0.8 fm^{-1}) = -2.53 fm^2 ;
C_{2}(\mu, p_F = 0.8 fm^{-1}) = 1.16 fm^4
\end{eqnarray}
For the potential energy per particle we find
 $\frac{U(p_F}{A}$ = - 11.5 MeV (-9.7 MeV, - 8.1 MeV) at $p_F$ = 1.2 
$fm^{-1}(1.0fm^{-1},0.8fm^{-1})$ respectively.
These results look quite reasonable, although somewhat smaller than the values
 obtained in more traditional approaches. We stopped the calculations at
 $p_F = 1.36fm^{-1}$ which is rather close to the nuclear saturation point.
The approach should be modified in the several aspects in order to be applied
to the nuclear systems with the density higher then the normal nuclear one.  
First, the Fermi-momentum, being only ``marginally light scale'' already at
normal nuclear density  becomes closer to the generic ``heavy'' scale.
Consequently, the counting rules must be modified  and the value 
of the effective cutoff increased. At some critical density the value of cutoff  
can reach the point, where the other degrees of freedom such as vector mesons and 
$\Delta$-isobars are to be included explicitly. It brings in the other sources of 
uncertainties which are rather difficult to control.
Moreover, whereas at normal nuclear density and typical energy scale $Q\sim$ 100 MeV
the relativistic corrections enter at higher orders, at some critical density
the whole formalism should be formulated in a fully relativistic way. It considerably
complicates the practical use of EFT. The counting rules must be 
significantly modified to take into account the excitations both the Fermi and Dirac
 seas and the fact that the nucleon mass can no longer be considered as a genuine
``heavy'' scale. On the other hand, at densities much higher than the normal 
nuclear one the matter was shown to be a color superconductor \cite {Wi} where
quark and gluon degrees of freedom must be considered explicitly. When the density
 is lowered a color superconducting phase should match the EFT, formulated in terms
of baryons and mesons. Carring out this matching is a highly nontrivial problem,
requiring at least qualitative  understanding of the mechanism of hadronization at
high densities \cite{PeB}. Therefore some model of how confinement
 occurs at high densities is probably needed.

As already mentioned above, to be completely consistent, many-body forces and
 pion effects must be included in the EFT treatment. The contribution from the
 many-body forces is expected to be rather small. There are at least two  reasons for
 such an expectations. First, each additional interacting nucleon line will bring in 
the additional small factor $Q^3 \sim p_{F}^3$ in the expression for 
the binding energy of nuclear matter. Second, following the line of arguments
 suggested in \cite {Fr} the contributions of the N-nucleon forces can schematically
be written as $V_N \sim Q^{N}/\Lambda^{N-1}$ where $\Lambda$ is the generic
 large-mass QCD scale close to 1 GeV and Q is the typical low-energy scale taken to
 be about 0.1 GeV. Using these numbers one can see that the contribution of the 
three-body forces is order of magnitude smaller compared to the two body ones and 
that four-body interaction can safely be neglected. We stress, however, that taking into
 account the 3-nucleon forces may turn out important to get the saturation point
 close  to the empirical one. The expected smallness of the many-body forces
compared to the two-body ones is also confirmed by the results of several other
studies both for nuclear matter and for finite nuclei with $A > 3$ \cite {Ca}. 

Pion degrees of freedom, being another important ingredient of any nuclear EFT,
 seem to average to NLO effects, when the S-wave part of nuclear matter
 binding energy is being calculated. One notes that this conclusion is partly
 supported by the results, obtained in the phenomenologically very successful
Skyrme approach \cite {Sk,Fu2}. The Skyrme interaction has the pointlike form
 and does not contain the explicit pions. In some sense the Skyrme interaction can
be considered  as a parameterization of the G-matrix \cite {Va},
 written in the form closed to that we used.
Furthermore, as shown in \cite{Ly} adding explicit pions can lead  only to the
moderate changes of the effective couplings. This fact can qualitatively be
 explained by the partial cancelation from the pointlike interaction and
 one-pion-exchange interaction. The existence of such a cancelation was also pointed
out in \cite{Lu}. However, it is worth noting that the
relative contribution due the pionic effects may depend on the partial channel
 considered. For example, we expect that the pion degrees of freedom 
are more important in the P-wave part of nuclear potential energy that in the S-wave 
one. 

There are a few other important issues worth discussing. First, the calculations
must be self-consistent in a sense that both G-matrix and effective nucleon potential,
entering the BG equation should be determined at the same order
from the system of coupled equations
since effective nucleon potential is itself determined by the G-matrix.
 This is not done yet. Second
remark concerns the problem of removing of the off-shell ambiguities. EFT offers,
in principle, the potential possibility to avoid this difficulty order by order.
In practice, however, it may turn out that the cancelation of the unwanted
 contributions can be achieved only if all graphs, relevant at a given chiral
order are included. It may turn out very complicated technical task.
Third, the nuclear dressing of pions is to be considered. In the other words,
 pion self-energy must be computed to the order dictated by the counting rules.

In conclusion let us list the main ingredients which have to be included in 
any consistent nuclear EFT.

1. The local interaction terms as well as pion degrees of freedom must be
included, probably up to NLO in the S, P and D-waves.

2. Nucleon and pion self-energies should be computed at a given chiral order.

3. To explain the saturation the three-nucleon forces seem to be needed.

4. The relations between the in-medium effective couplings and those
arising in the vacuum 2-body and few-body calculations have to be established.

5. The reasonable way to generalize the nuclear matter EFT to treat the finite
nuclei should be found.

6. All that must be supplied with the consistent formulation of the in-medium
counting rules. This is probably a crucial issue for future studies.

As seen from this list  we are at the beginning of the road. Many things must
be done to formulate the entirely consistent and yet practical approach.
However, the results already obtained on this way give a good reason to hope
on a further progress in applying of the EFT methods to nuclear physics.    
\newpage

\bf {REFERENCES}

\end{document}